\documentclass{ws-ijmpe}
\usepackage[super,compress]{cite}
\begin{document}

\markboth{N. Ashok, A. Joseph}{Alpha and cluster decay in Pt isotopes}

\catchline{}{}{}{}{}

\title{A systematic study of alpha and cluster decay in Platinum isotopes}

\author{Nithu Ashok, Antony Joseph}

\address{Department of Physics, University of Calicut, Kerala, India.\\
nithu.ashok@gmail.com}

\maketitle

\begin{history}
\received{Day Month Year}
\revised{Day Month Year}
\end{history}

\begin{abstract}
The feasibility of alpha and cluster decay from Pt isotopes has been investigated within the framework of Skyrme Hartree-Fock-Bogoliubov theory.  Calculation has been carried out for various Skyrme forces.  Harmonic oscillator and transformed harmonic oscillator basis are used to solve HFB equations.  The role played by shell closure is withal analysed.  Half-lives are estimated with the help of Universal Decay Law (UDL).  Geiger-Nuttel plots are also plotted and successfully preserves its linear nature.
\end{abstract}


\ccode{PACS numbers: 21.10.Dr, 21.60.Jz, 23.70.+j}


\section{Introduction}

Cluster radioactivity is an exotic decay, in which the mass of emitted fragment lies between the alpha particle and the lightest fission fragment.  No neutrons are emitted in this process\cite{ref1}.  The phenomena of cluster radioactivity were first predicted by Sandulescu et al. in 1980\cite{sand}, which was later experimentally observed by Rose and Jones, in 1984\cite{rose}.  They had observed the emission of $^{14}\textrm{C}$ cluster from $^{223}\textrm{Ra}$.  Experimental confirmation of the same has been carried out by several other groups in subsequent years\cite{alek,gal,kut}.    Several other decay modes like $^{20}\textrm{O}$, $^{22,24-26}\textrm{Ne}$, $^{28-30}\textrm{Mg}$ and $^{32,34}\textrm{Si}$ have been experimentally observed so far\cite{barw,bon1,bon2}.  Cluster radioactivity is a rare cold nuclear phenomenon which is theoretically explained with the aid of quantum mechanical fragmentation theory (QMFT).

Different theoretical models are used to explore this exotic decay.  One is the Unified Fission Model (UFM)\cite{ufm} in which the valence nucleons of the parent nucleus condense to form the cluster, through continuous dynamical changes, which finally emerges out by barrier penetration.  The other one is the Preformed Cluster Model (PCM)\cite{pcm}, in which cluster is assumed to be pre-formed inside the nucleus and it will undergo tunnelling through the potential barrier. The main difference lies in the fact that, UFM is based on the assumption that cluster will surely be emitted and preformation probability is always taken to be one.  But in the case of PCM, we have to calculate it explicitly.  In the present work, our result has been compared with Effective Liquid Drop Model (ELDM) which is a fission-like model.  Theoretical and experimental studies on cluster radioactivity have been carried out in recent years by various groups throughout the world.  Different studies show that this phenomenon occurs in those regions where daughter nuclei should either be doubly magic or in its vicinity.  In view of this observation, cluster radioactivity falls into two regions, trans-tin and trans-lead.  Our study mainly falls in the trans-tin region.

In our previous works, we have studied the feasibility of this exotic decay in tungsten (W)\cite{w} and osmium (Os)\cite{os}  isotopes.  Nuclei in the rare earth region, i.e., in the mass range $150<A<190$ are good candidates for the investigation of this exotic decay.  In the present work, we have extended our survey on cluster decay to the Pt(Z=78) isotopes.  Moreover, Pt which is a transitional nucleus is found to be deformed in its ground state and can be expected to be unstable against some heavy decay modes\cite{pt}.  Several studies on alpha decay of Pt isotopes has been carried out in recent decades, both theoretically\cite{rt1,rt2,rt3} and experimentally\cite{re1,exp1,exp2,exp3}.

The paper is organised as follows.  In section 2, we have discussed briefly the theoretical formalism,  Hartree-Fock-Bogoliubov theory, which is employed for the present investigation.  In section 3, we have shown the details of the calculation.  
In section 3, we have presented the Results and discussion of the work, where we have shown the sensitivity of different Skyrme forces in predicting the half-lives of alpha and cluster radioactivity in Pt isotopes.  Final summary and conclusion are given in section 5.  

\section{Theoretical framework}
The present study has been carried out with the help of the Hartree-Fock-Bogoliubov (HFB) theory.  HFB theory is a combination of both Hartree-Fock (HF) and BCS theory.  In HFB theory, self-consistent field and pairing field are given equal importance\cite{ring}.  More details of the theory can be found in our previous paper \cite{w}.  HFB equations are solved using axially deformed harmonic oscillator (HO) and transformed harmonic oscillator (THO) basis \cite{hfbtho}.

In the mean-field part, zero range Skyrme effective interaction is used.  There exist a wide range of Skyrme forces in literature.  Among those, we have selected six parametrizations which are highly efficient in reproducing the ground state properties.  They are SIII\cite{siii}, SkP\cite{skp},  SLy5\cite{sly5},  SkM*\cite
{skm}, UNEDF0\cite{une0} and UNEDF1\cite{une1}.  The selected Skyrme parameterization includes the classic one which is SIII, and also the very recent UNEDFs.  SIII is designed for predicting binding energies and densities, and single particle energy levels around Fermi level.  SkP is designed especially for HFB equations with effective mass $m^{*}/m=1$ to include the effect of pairing.  SLy5 is designed for the infinite nuclear matter with the inclusion of $J^{2}$ term.  SkM* which is the modified version of SkM, has been designed to adjust the fission barriers of  $^{240}\textrm{Pu}$ and has also been optimized for large deformation.  UNEDF's are aimed in predicting of spectroscopic properties of nuclei based on nuclear energy density functional.  UNEDF0 predicts experimental masses, radii, and deformations fairly well whereas UNEDF1, as an extension, especially aims in fission and fusion studies.
 
In the pairing part, the density dependent delta interaction(DDDI)\cite{dddi, dddi1} in its mixed form is used.
The mixed variant of density dependent delta interaction(DDDI) is given by\cite{mix},
\begin{equation}
  V^{n/p}_{\delta }(\vec{r_{1}},\vec{r_{2}})=V_{0}^{n/p}[1-\frac{1}{2}(\frac{\rho(\vec{r_{1}} +\vec{r_{2}})}{\rho _{0}} )^{\alpha }]\delta (\vec{r_{1}}-\vec{r_{2}})
\end{equation}
where the saturation density\cite{sat} $\rho _{0}$=0.16 fm$^{-3}$ and $\alpha $=1.

Several semi-empirical formulae have been developed to predict the half-lives of various decay modes.  Here we have adopted the Universal Decay Law(UDL)\cite{udl,udl1} which has been deduced from WKB approximations, with some modifications.  It is given by, 
\begin{equation}
log_{10}T_{1/2}=aZ_{c}Z_{d}\sqrt{\frac{A}{Q}}+b\sqrt{AZ_{c}Z_{d}(A_{c}^{1/3}+A_{d}^{1/3})}+c
\end{equation}
where the constants are a=0.4314, b=-0.4087 and c=-25.7725.\\ $Z_{c}$, $Z_{d}$ are the atomic number of cluster and daughter nuclei , $A_{c}$, $A_{d}$ are the mass number of cluster and daughter nuclei and 
\begin{equation}
 A=\frac{A_{c}A_{d}}{A_{c}+A_{d}}
  \end{equation}
Here Q is the Q-value of the decay.

Moreover, our study mainly concentrates on those decays having the half-lives in the experimentally measurable range $(T_{1/2}<10^{30}s)$.


\section{Result and Discussion}
The present work is devoted to the study of alpha decay and cluster radioactivity in Pt isotopes in the neutron-deficient region.  As a continuation of our previous works, we have made an attempt to predict the feasibility of these decay modes in the even-even Pt isotopes between 2p-drip line and the beta stability line within Skyrme HFB framework.

\begin{figure}[h!]
\centerline{\includegraphics[width=4in]{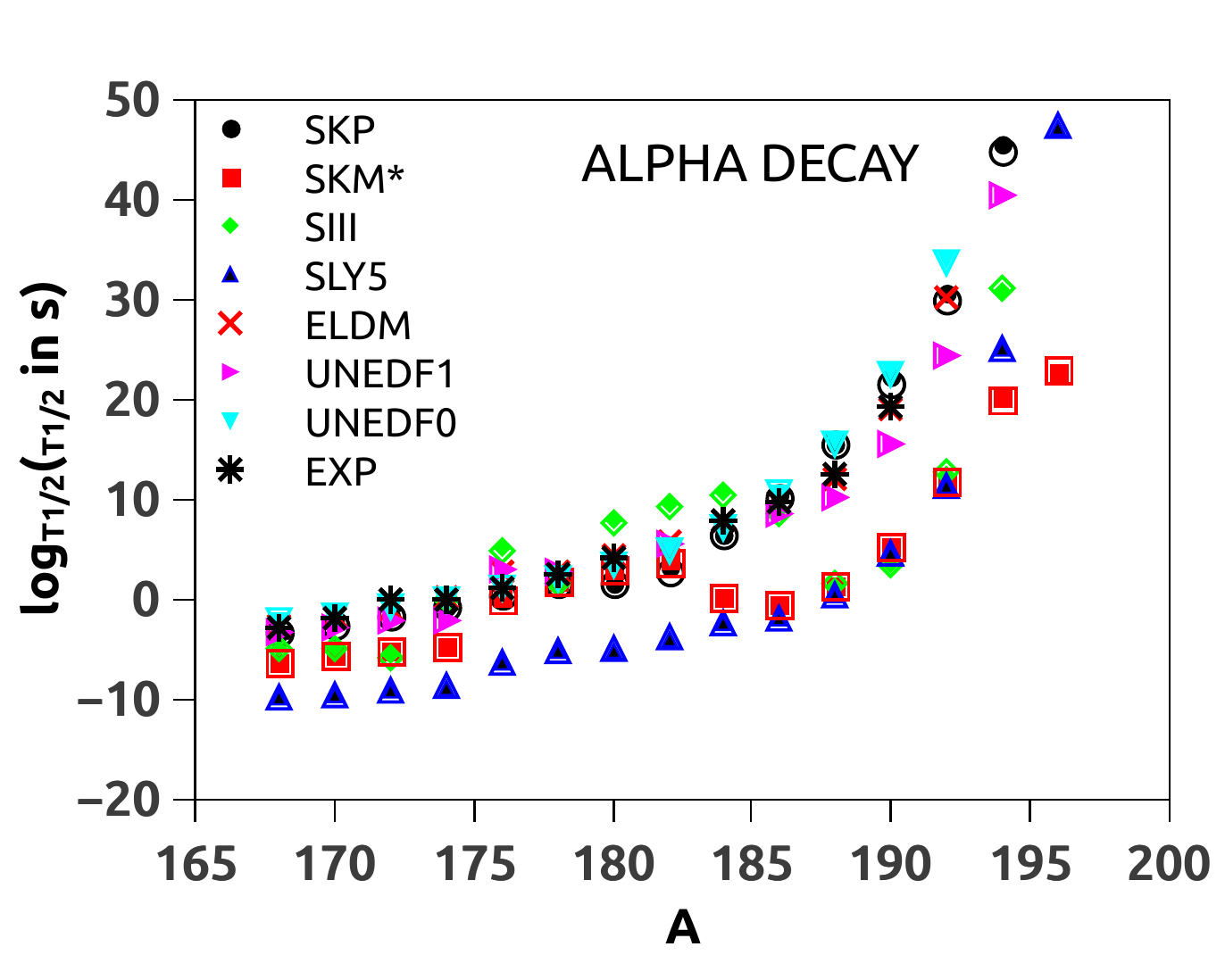}}
\vspace*{8pt}
\caption{Plot showing the logarithmic value of half-life $(T_{1/2}$ in s) against the mass number of parent (A) nuclei corresponding to alpha decay for HO(solid) and THO(open) basis\protect\label{pt-alpha}.}
\end{figure}

Alpha decay is one of the prominent decay modes which is exhibited by the atomic nucleus.  In the first part of the study, we have tried to predict the feasibility of alpha decay in Pt isotopes.  The half-lives are calculated using the UDL given by the equation (2).  From this equation, it is clear that the half-lives depend on the Q-value of the reaction.  
$Q_{\alpha }$-values are calculated from binding energies using the relation,
\begin{equation}
Q_{\alpha} (N,78)=B(N-2,76)+B(2,2)-B(N,78)
\end{equation}
where, B(N,78) and B(N-2,76) are the binding energies of the parent (Pt) and the daughter nucleus (Os).  B(2,2) is the binding energy of $_{2}^{4}\textrm{He}$ nucleus (28.296 MeV), which is taken from AME 2012 \cite{ame}.  Table \ref{tab:pt-alp} shows the predicted $Q_{\alpha }$-values.  The microscopic values are compared with the phenomenological ELDM values as well as the experimental values \cite{exp1,exp2,exp3}.  Predicted half-lives are depicted in fig. \ref{pt-alpha} from which we can see that SLY5 values underestimate the experimental values highly compared to other Skyrme forces.  UNEDF's predicts the half-lives close to experimetal ones.  In table \ref{tab:pt-sd}, we have shown the standard deviation of the predicted values in the case of each Skyrme forces.  UNEDF values shows less deviation in predicting alpha-decay half-lives compared to others.   

In the second part of the study, we have extended our calculation to predict the clusters which are likely to be emitted from Pt isotopes.  Here also the half-lives are computed using the equation (2).  Also the Q-values for various clusters are given by,
$^{8}\textrm{Be}$:
\begin{equation}
Q(N,78)=B(N-4,74)+B(4,4)-B(N,78)
\end{equation}
$^{12}\textrm{C}$:
\begin{equation}
Q(N,78)=B(N-6,72)+B(6,6)-B(N,78)
\end{equation}
$^{16}\textrm{O}$:
\begin{equation}
Q(N,78)=B(N-8,70)+B(8,8)-B(N,78)
\end{equation}
$^{20}\textrm{Ne}$:
\begin{equation}
Q(N,78)=B(N-10,68)+B(10,10)-B(N,78)
\end{equation}
$^{24}\textrm{Mg}$:
\begin{equation}
Q(N,78)=B(N-12,66)+B(12,12)-B(N,78)
\end{equation}
where, B(N-4,74), B(N-6,72), B(N-8,70), B(N-10,68), B(N-12,66) are the binding energies of the corresponding daughter nuclei (W, Hf, Yb, Er and Dy)
and B(4,4), B(6,6), B(8,8), B(10,10) and B(12,12) 
are the binding energies of the emitted clusters $^{8}\textrm{Be}$, $^{12}\textrm{C}$, $^{16}\textrm{O}$, $^{20}\textrm{Ne}$ and $^{24}\textrm{Mg}$ respectively.  
As in the case of $\alpha$-decay, we have shown the Q-values of the emitted clusters in table \ref{tab:pt-clus}.  Due to the lack of availability of experimental values, we have compared them with ELDM values.  We observed a good agreement among the values with a small discrepency in some cases.  

\begin{figure}[t]
\centerline{\includegraphics[width=6in]{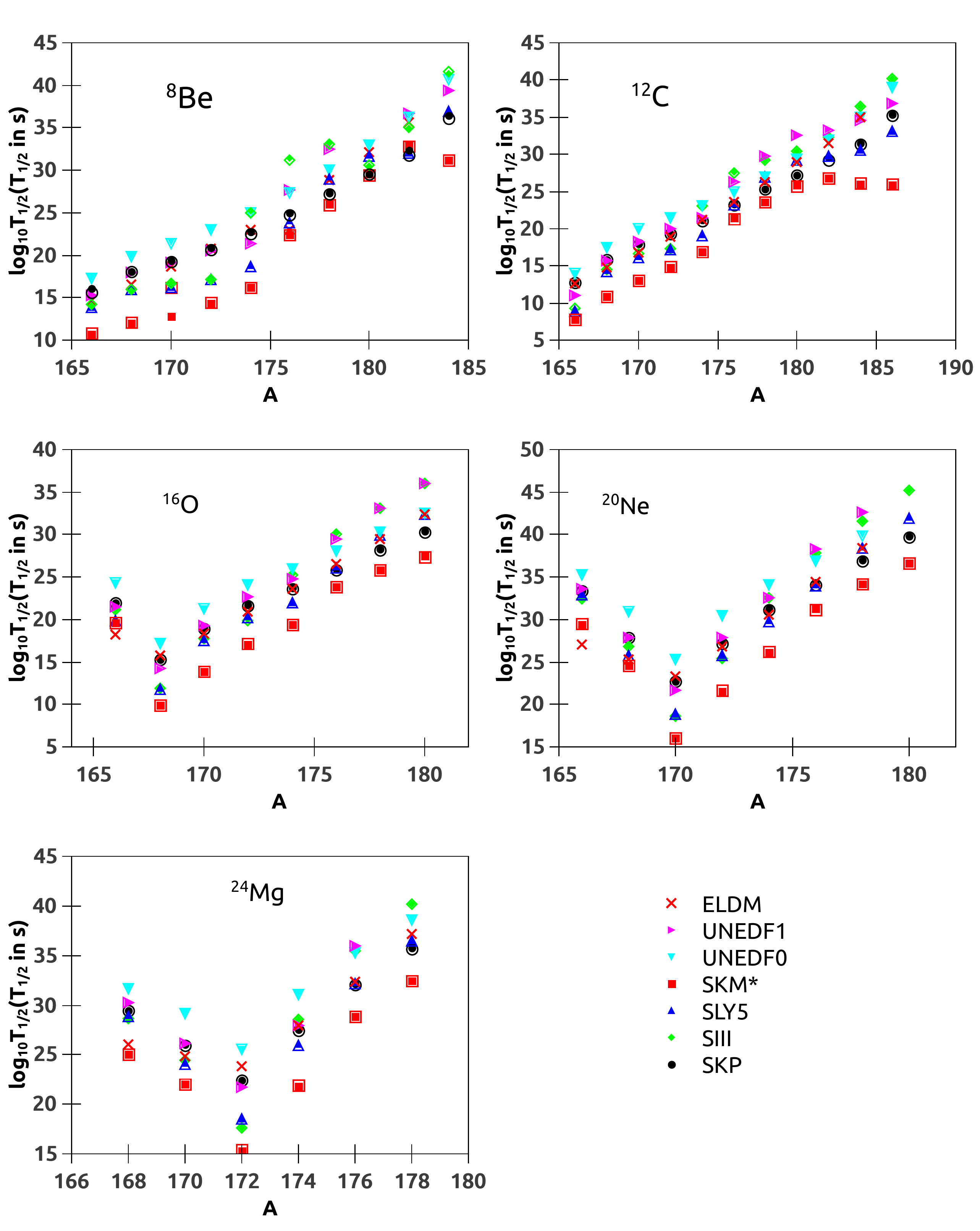}}
\vspace*{8pt}\caption{Plots showing logarithmic value of half-life $(T_{1/2}$ in sec) against mass number of parent (A) nuclei, corresponding to different cluster decay modes for HO(solid) and THO(open) basis.\protect\label{pt-clus}}
\end{figure}

From the earlier works, it is clear that the binding energy of an isotope depends on the type of Skyrme force employed.  As the Q-values are computed from binding energies, its effect is reflected in the values given in the table.  Using these Q-values, we can estimate the half-lives using the equation (2).  Computed half-lives are plotted in fig. \ref{pt-clus}.  With ELDM values taken as reference, we can say that SKP and UNEDF1 values matches well with them, while SKM* values underestimates.  In the case of Pt isotopes, we have observed five clusters which are having the half-lives falling in the experimentally measurable range.  A comparison with the previous works, shows that as the atomic number of the parent nuclei increases, more massive clusters will be emitted.  It is also observed that the rate of cluster decay depends on the neutron number of the parent.  As the neutron number increases the possibility of cluster emission diminishes.  This phenomena is dominant in the neutron deficient region of the isotopic chain.

It is well known that shell closure have a major role in the phenomena of cluster radioactivity.  From the fig. \ref{pt-clus}, we have observed that the half-lives are minimum for those decays which are having the daughter nuclei with magic neutron number.  This is visbile only for $^{16}\textrm{O}$, $^{20}\textrm{Ne}$ and $^{24}\textrm{Mg}$ decays only and their magic daughters are $^{168}\textrm{Yb}$, $^{150}\textrm{Er}$ and $^{148}\textrm{Dy}$ respectively with neutron number N=82.  For $^{8}\textrm{Be}$ and $^{12}\textrm{C}$, the magic daughter nuclei lies outside the selected region.  So it is not shown here.

\begin{figure}[t]
\centerline{\includegraphics[width=6in]{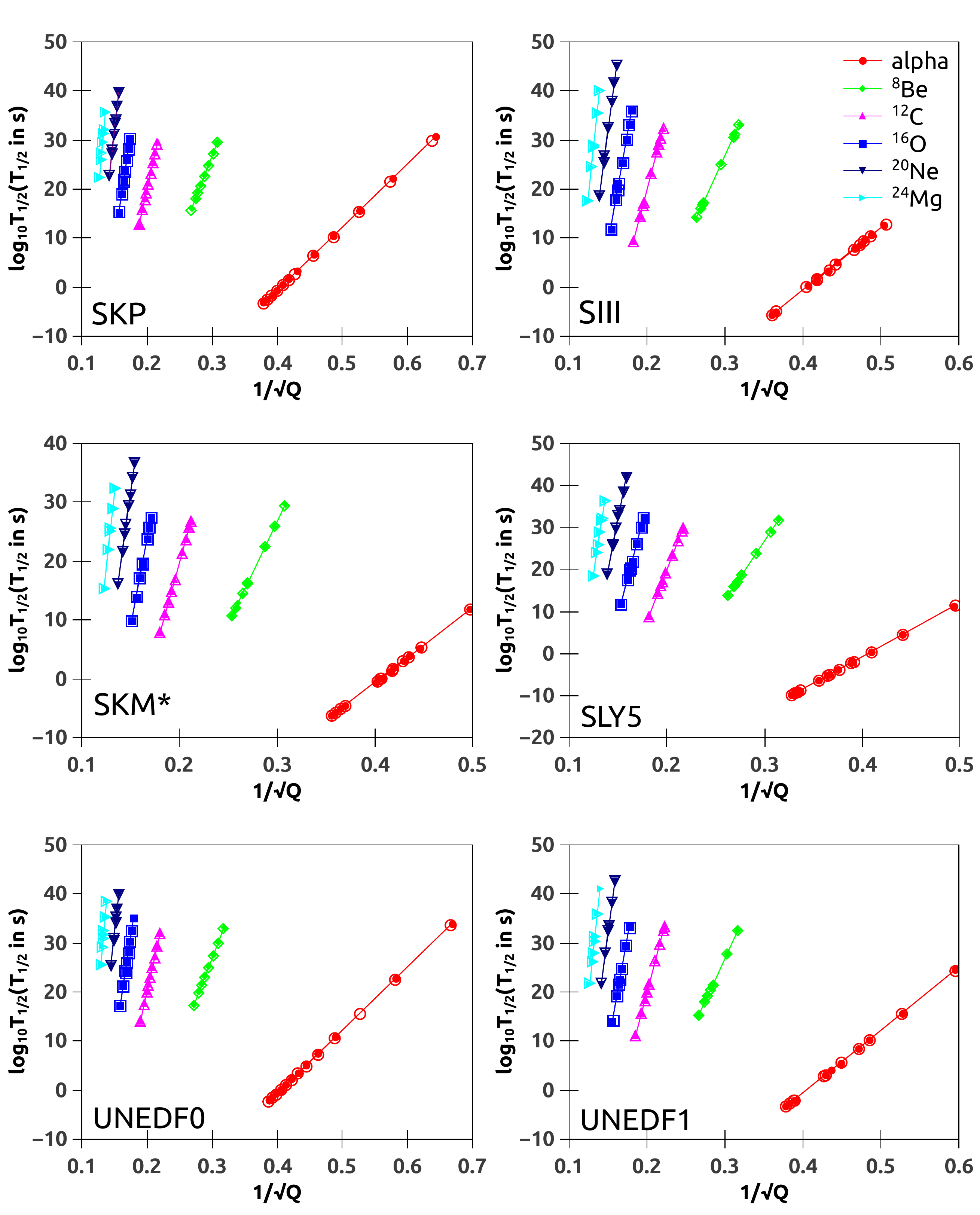}}
\vspace*{8pt}
\caption{Geiger-Nuttal plots of different cluster decay modes for HO(solid) and THO(open) basis corresponding to different Skyrme forces.\protect\label{pt-gnp}}
\end{figure}

Geiger-Nuttel (GN) plots are shown in fig. \ref{pt-gnp}.  In the case of all the selected Skyrme forces, the linear nature of the GN plot has been sucessfully reproduced.  By least square fitting, we were able to identify the slope as well as the intrcept of the GN plots.  As in the previous results, here also we observed that the increase of the slope shows the emission of the massive cluster.  We have computed the slopes and intercept for each emitted cluster and for each Skyrme interaction.  The calculated values are tabulated in table \ref{tab:pt-slope}.  All the Skyrme forces have similar slope and intercept. 

\begin{table}[ht]
\caption{\label{tab:pt-alp}Q-values of alpha decay in even-even Pt isotopes calculated with Skyrme HFB equations solved using HO(top) and THO(bottom) basis along with ELDM and available experimental values.}
{\resizebox{13cm}{!}{
 \begin{tabular}{ccccccccc} \hline
\multicolumn{1}{c}{Alpha decay} & \multicolumn{7}{c}{Q value (MeV)}\\ \cline{2-9}
& SIII & SKP & SkM* & SLy5 & UNEDF0 & UNEDF1 & ELDM  & exp  \\ \hline
$^{168}\textrm{Pt}\rightarrow \alpha +^{164}\textrm{Os}$&7.5095&6.9677&7.9371&9.3412&6.6857&7.0041&6.9851&-\\
&7.5017&6.9957&7.9191&9.3417&6.7063&6.6421&&\\
$^{170}\textrm{Pt}\rightarrow \alpha +^{166}\textrm{Os}$&7.4979&6.7617&7.7329&9.0282&6.5111&6.8369&6.7071&-\\
&7.4882&6.7752&7.7270&9.1571&6.5195&6.6326&&\\
$^{172}\textrm{Pt}\rightarrow \alpha +^{168}\textrm{Os}$&7.7265&6.5345&7.5196&9.0282&6.3062&6.6312&6.4651&-\\
&7.7374&6.5527&7.5182&8.9980&6.3246&5.4551&&\\
$^{174}\textrm{Pt}\rightarrow \alpha +^{170}\textrm{Os}$&6.0724&6.2828&7.3505&8.8437&6.0872&6.6242&6.1831&6.03\\
&6.1191&6.2924&7.3454&8.8209&6.1091&5.4946&&\\
$^{176}\textrm{Pt}\rightarrow \alpha +^{172}\textrm{Os}$&5.0881&6.0046&6.0777&7.9131&5.8545&5.4481&5.8851&5.74\\
&5.1407&6.0159&6.1019&7.9087&5.8805&-&&\\
$^{178}\textrm{Pt}\rightarrow \alpha +^{174}\textrm{Os}$&5.7452&5.7459&5.7023&7.5347&5.6046&-&5.5731&5.44\\
&5.7283&5.7598&5.7227&7.5390&5.6363&4.9739&&\\
$^{180}\textrm{Pt}\rightarrow \alpha +^{176}\textrm{Os}$&4.6060&5.7321&5.4355&7.4556&5.3591&5.2652&5.2371&5.14\\
&4.6371&5.7503&5.4566&7.4421&5.3764&-&&\\
$^{182}\textrm{Pt}\rightarrow \alpha +^{178}\textrm{Os}$&4.3582&5.3930&5.2888&7.0965&5.0605&4.9783&4.9511&4.84\\
&4.3835&5.5048&5.3127&7.0684&5.0841&4.9739&&\\
$^{184}\textrm{Pt}\rightarrow \alpha +^{180}\textrm{Os}$&4.2003&4.8019&6.0484&6.6459&4.6734&-&4.5981&4.50\\
&4.2222&4.8319&6.0461&6.6364&4.6906&-&&\\
$^{186}\textrm{Pt}\rightarrow \alpha +^{182}\textrm{Os}$&4.4666&4.2332&6.2034&6.5487&4.1855&4.5068&4.3201&4.23\\
&4.4775&4.2450&6.1889&6.5475&4.1979&4.5043&&\\
$^{188}\textrm{Pt}\rightarrow \alpha +^{184}\textrm{Os}$&5.7406&3.6178&5.7841&5.9810&6.0632&4.2614&4.0027&3.93\\
&5.7241&3.6245&5.7536&5.9732&3.6108&2.8270&&\\
$^{190}\textrm{Pt}\rightarrow \alpha +^{186}\textrm{Os}$&5.3471&2.9960&5.0134&5.1340&2.9514&3.6010&3.2525&3.18\\
&5.3184&3.0516&4.9977&5.1316&2.9712&3.6045&&\\
$^{192}\textrm{Pt}\rightarrow \alpha +^{188}\textrm{Os}$&3.9417&2.4168&4.0453&4.0952&2.2386&2.8239&2.4224&2.6\\
&3.9107&2.4662&4.0434&4.0873&2.2538&2.8269&&\\

\hline
\end{tabular}}}
\end{table}


\begin{table*}[h]
\caption{\label{tab:pt-sd}Comparison of standard deviation of alpha decay half-lives of Pt isotopes calculated for different Skyrme forces }
\centering
   {\begin{tabular}{ccccccc} \hline
& SKP & SLY5&SIII & SKM* & UNEDF0&UNEDF1\\ \hline
HO&1.6367&8.0337&1.9835&2.9102&0.7660&1.4106\\
THO&1.5632&8.0155&1.9333&2.9088&0.7127&1.5111\\
\hline
  
  \end{tabular} }
\end{table*} 


\begin{table}[!]\footnotesize
\caption{\label{tab:pt-clus}Same as Table 1, but for various clusters}
\centering
\renewcommand{\arraystretch}{0.8}
  {\begin{tabular}{@{}*{9}{c}@{}} \hline
\multicolumn{1}{c}{Cluster decay} & \multicolumn{7}{c}{Q value (MeV)}\\ \cline{2-8}  
& SIII & SKP & SkM* & SLy5 & UNEDF0 & UNEDF1 & ELDM   \\ \hline
$^{168}\textrm{Pt}\rightarrow ^{8}\textrm{Be} +^{158}\textrm{W}$&13.8943&13.2826&15.1167&13.8505&12.7603&13.3022&13.3783\\
&13.8817&13.2698&15.0931&13.8843&12.7904&13.3093&\\
$^{170}\textrm{Pt}\rightarrow ^{8}\textrm{Be} +^{160}\textrm{W}$&13.6507&12.9141&14.8164&13.7675&12.3563&12.9551&12.7533\\
&13.6301&12.9405&13.7924&13.7921&12.3886&12.9713&\\
$^{172}\textrm{Pt}\rightarrow ^{8}\textrm{Be} +^{162}\textrm{W}$&13.4963&12.5140&14.3053&13.4782&11.9556&12.5739&12.1893\\
&13.5003&12.5487&14.2876&13.5086&11.9830&12.5992&\\
$^{174}\textrm{Pt}\rightarrow ^{8}\textrm{Be} +^{164}\textrm{W}$&11.4772&12.0364&13.7439&13.0293&11.4798&12.3392&11.6283\\
&11.5213&12.0721&13.7315&13.0483&11.5123&12.3575&\\
$^{176}\textrm{Pt}\rightarrow ^{8}\textrm{Be} +^{166}\textrm{W}$&10.2062&11.5136&12.0880&11.7145&10.9630&10.9194&11.0173\\
&10.2574&11.5656&12.1034&11.7592&11.0016&10.9349&\\
$^{178}\textrm{Pt}\rightarrow ^{8}\textrm{Be} +^{168}\textrm{W}$&9.8748&10.9932&11.2687&10.6084&10.4305&9.9887&10.3513\\
&9.8988&11.0128&11.2845&10.6598&10.4651&10.0089&\\
$^{180}\textrm{Pt}\rightarrow ^{8}\textrm{Be} +^{170}\textrm{W}$&10.3562&10.5249&10.5363&10.1057&9.8851&-&9.7193\\
&10.3296&10.5282&10.5468&10.1325&9.9162&-&\\ \\
$^{168}\textrm{Pt}\rightarrow ^{12}\textrm{C} +^{156}\textrm{Hf}$&27.4275&26.8167&29.1631&27.4953&26.0535&26.8678&26.8100\\
&27.4153&26.7925&29.1387&27.5429&26.0803&26.8670&\\
$^{170}\textrm{Pt}\rightarrow ^{12}\textrm{C} +^{158}\textrm{Hf}$&26.4482&25.8387&28.0825&26.5873&25.0066&25.7173&25.7980\\
&26.4338&25.8663&28.0595&26.6225&25.0360&25.7283&\\
$^{172}\textrm{Pt}\rightarrow ^{12}\textrm{C} +^{160}\textrm{Hf}$&26.0786&25.2146&27.1841&26.1268&24.3644&24.9612&24.8340\\
&26.0806&25.2601&27.1582&26.1557&24.4079&24.9790&\\
$^{174}\textrm{Pt}\rightarrow ^{12}\textrm{C} +^{162}\textrm{Hf}$&23.6871&24.4555&26.2305&25.2748&23.7177&24.3177&23.8510\\
&23.7293&24.5007&26.2174&25.2957&23.7657&24.3398&\\
$^{176}\textrm{Pt}\rightarrow ^{12}\textrm{C} +^{164}\textrm{Hf}$&22.1017&23.6228&24.3359&23.5920&23.0132&22.5277&22.8840\\
&22.1461&23.6573&24.3547&23.6459&23.0557&22.5612&\\
$^{178}\textrm{Pt}\rightarrow ^{12}\textrm{C} +^{166}\textrm{Hf}$&21.5515&22.8486&23.4794&22.2496&22.2457&21.3546&21.8620\\
&21.5744&22.8731&23.4920&22.3088&22.2905&21.3867&\\
$^{180}\textrm{Pt}\rightarrow ^{12}\textrm{C} +^{168}\textrm{Hf}$&21.1318&22.1688&22.6619&21.4802&21.4602&20.4886&20.9250\\
&21.1445&22.1844&22.6697&21.5253&21.4932&20.5088&\\
$^{182}\textrm{Pt}\rightarrow ^{12}\textrm{C} +^{170}\textrm{Hf}$&20.5519&21.3884&22.2981&21.3192&20.6322&20.2435&20.0860\\
&20.5421&21.4945&22.3013&21.3374&20.6745&20.2568&\\ \\
$^{168}\textrm{Pt}\rightarrow ^{16}\textrm{O} +^{152}\textrm{Yb}$&42.6291&41.1181&43.8462&42.6181&39.5458&41.2947&39.9970\\
&42.5866&40.5989&43.8134&42.6509&39.5857&41.2530&\\
$^{170}\textrm{Pt}\rightarrow ^{16}\textrm{O} +^{154}\textrm{Yb}$&39.2229&39.1071&41.3485&39.2865&37.3761&38.4226&38.3640\\
&39.2106&38.5879&41.3273&39.3334&37.3977&38.4257&\\
$^{172}\textrm{Pt}\rightarrow ^{16}\textrm{O} +^{156}\textrm{Yb}$&38.0085&37.5935&39.4978&37.8204&35.9506&36.6511&36.8980\\
&38.0221&37.1739&39.4734&37.8571&35.9885&36.6634&\\
$^{174}\textrm{Pt}\rightarrow ^{16}\textrm{O} +^{158}\textrm{Yb}$&35.3539&36.4367&38.2173&36.9080&35.0079&35.5997&35.4290\\
&35.3595&36.1677&38.1962&36.9210&35.0625&35.6150&\\
$^{176}\textrm{Pt}\rightarrow ^{16}\textrm{O} +^{160}\textrm{Yb}$&33.1992&35.1748&35.9715&34.8878&34.0361&33.4533&33.9680\\
&33.2378&35.0667&35.9904&34.9310&34.0872&33.4738&\\
$^{178}\textrm{Pt}\rightarrow ^{16}\textrm{O} +^{162}\textrm{Yb}$&31.9616&33.9330&35.0054&33.1724&33.0492&31.9147&32.567\\
&31.9792&33.9896&35.0207&33.2259&33.0935&31.9428&\\
\\
$^{168}\textrm{Pt}\rightarrow ^{16}\textrm{Ne} +^{148}\textrm{Er}$ &47.2589&47.1890&48.4482&47.7259&45.1348&46.6928&47.4609\\
&47.2533&46.6797&48.4356&47.7864&45.1655&46.6841&\\
$^{170}\textrm{Pt}\rightarrow ^{16}\textrm{Ne} +^{150}\textrm{Er}$ &51.8056&49.9402&53.3843&51.6341&47.9225&49.9919&48.5679\\
&51.7622&49.4091&53.3576&51.6626&47.9655&49.9697&\\
$^{172}\textrm{Pt}\rightarrow ^{16}\textrm{Ne} +^{152}\textrm{Er}$ &47.8083&47.2965&49.9549&47.6138&45.2075&46.5115&46.4389\\
&47.8137&46.8730&49.9298&47.6643&45.2427&46.5266&\\
$^{174}\textrm{Pt}\rightarrow ^{16}\textrm{Ne} +^{154}\textrm{Er}$ &44.1276&45.1223&47.3675&45.4749&43.3592&44.1470&44.3289\\
&44.1726&44.8489&47.3469&45.4976&43.4102&44.1566&\\
$^{176}\textrm{Pt}\rightarrow ^{16}\textrm{Ne} +^{156}\textrm{Er}$ &41.6704&43.4988&44.7287&43.3912&42.0585&41.4709&42.3179\\
&41.7075&43.3828&44.7411&43.4209&42.1141&41.4929&\\
$^{178}\textrm{Pt}\rightarrow ^{16}\textrm{Ne} +^{158}\textrm{Er}$ &40.0433&42.0026&43.2369&41.3226&40.7041&39.6015&40.3489\\
&40.0550&42.0630&43.2497&41.3603&40.7580&39.6422&\\ \\

$^{168}\textrm{Pt}\rightarrow ^{24}\textrm{Mg} +^{144}\textrm{Dy}$&58.2312&58.2251&60.3335&58.0293&56.4468&57.2736&59.4436\\
&58.2292&57.7207&60.3208&58.0965&56.4949&57.2727&\\
$^{170}\textrm{Pt}\rightarrow ^{24}\textrm{Mg} +^{146}\textrm{Dy}$&60.5688&60.1970&62.0858&60.7839&57.7422&59.5464&60.1836\\
&60.5643&59.6774&62.0738&60.8262&57.8136&59.5416&\\
$^{172}\textrm{Pt}\rightarrow ^{24}\textrm{Mg} +^{148}\textrm{Dy}$&64.8511&62.1707&66.4221&64.2074&59.8010&62.1856&60.6896\\
&64.8403&61.7410&66.3836&64.2409&59.8506&62.1729&\\
$^{174}\textrm{Pt}\rightarrow ^{24}\textrm{Mg} +^{150}\textrm{Dy}$&57.9479&58.9026&61.9662&59.4237&56.5463&58.3230&57.9246\\
&57.9942&58.6273&61.9387&59.4552&56.5923&58.3354&\\
$^{176}\textrm{Pt}\rightarrow ^{24}\textrm{Mg} +^{152}\textrm{Dy}$&54.1181&56.0758&57.6819&55.8244&54.2087&53.8786&55.1176\\
&54.1682&55.9572&57.6922&55.8736&54.2738&53.8980&\\
$^{178}\textrm{Pt}\rightarrow ^{24}\textrm{Mg} +^{154}\textrm{Dy}$&51.7344&53.9332&55.6472&53.5544&52.4739&51.3033&52.3306\\
&51.7582&53.9926&55.6532&53.5857&52.5367&-&\\ 

\hline
\end{tabular}}
\end{table}

\begin{table}
\caption{\label{tab:pt-slope}Slopes and intercepts of even-even Pt isotopes calculated for different Skyrme forces using HO(top) and THO(bottom) basis }
\centering
\resizebox{\textwidth}{!}{ 
\setlength{\tabcolsep}{2.2pt}
   \begin{tabular}{@{}*{13}{ccc}@{}} \hline
 \multicolumn{1}{c}{Skyrme}& \multicolumn{2}{c}{Alpha}& \multicolumn{2}{c}{Be}&\multicolumn{2}{c}{C}&\multicolumn{2}{c}{O}&\multicolumn{2}{c}{Ne}&\multicolumn{2}{c}{Mg}\\
force & Slope & Intercept & Slope & Intercept & Slope & Intercept & Slope & Intercept & Slope & Intercept & Slope & Intercept \\ \hline

\hline
SKP&128.198&-51.815&341.170&-75.589&594.985&-99.103&874.118&-121.739&1182.924&-145.292&1502.437&-168.592 \\
&128.164&-51.802&341.286&-75.626&594.561&-99.021&873.755&-121.684&1182.939&-145.435&1503.212&-168.696 \\
SLY5&126.993&-51.476&344.951&-76.706&601.635&-100.475&885.710&-123.688&1195.328&-147.306&1521.823&-171.163\\
&126.995&-51.476&344.878&-76.688&601.508&-100.453&885.611&-123.674&1195.273&-147.300&1521.969&-171.183 \\
SIII&127.470&-51.536&346.225&-77.027&604.065&-100.913&892.051&-124.677&1198.884&147.789&1511.682&-169.807\\
&127.433&-51.521&346.144&-77.004&603.950&-100.890&891.921&-124.655&1198.797&-147.776&1511.719&-169.813 \\
SKM*&126.690&-51.272&345.174&-76.831&599.912&-100.229&883.933&-123.513&1193.222&-147.108&1509.444&-169.646\\
&126.664&-51.261&344.881&-76.741&599.801&-100.207&883.725&-123.478&1193.109&-147.091&1509.439&-169.645 \\
UNEDF0&128.445&-51.928&342.563&-75.920&596.366&-99.301&874.512&-121.694&1189.337&-146.265&1485.281&166.209\\
&128.257&-51.832&342.492&-75.903&596.237&-99.279&874.978&-121.775&1189.256&-146.256&1485.286&-166.214& \\
UNEDF1&127.658&-51.580&345.037&-76.673&603.206&-100.709&888.587&-124.071&1199.454&-147.825&1508.678&-169.371\\
&127.655&-51.578&344.998&-76.663&603.102&-100.689&888.391&-124.039&1199.341&-147.808&1534.976&-172.800\\
\hline   
   \end{tabular} }
   \end{table}  

\section{Conclusion}
The systematic study of the phenomena of alpha decay and cluster radioactivity has been carried out with the aid of Skyrme Hartree-Fock-Bogoliubov theory.  Calculations have been done with the help of harmonic and transformed harmonic oscillator basis.  Six different Skyrme parametrizations have been used for the study.  The use of different oscillator basis shows only a very small difference in the order of a few keV in their half-lives.  The calculated standard deviation shows that UNEDF parametrizations show less deviation in predicting the half-lives of alpha decay compared to other Skyrme parametrizations.  We have observed the emission of clusters like $^{8}\textrm{Be}$, $^{12}\textrm{C}$, $^{16}\textrm{O}$, $^{20}\textrm{Ne}$ and $^{24}\textrm{Mg}$.  Compared to previous studies, it can be concluded that as the atomic number of parent nuclei increases, we can expect the emission of massive clusters.  It was also observed that the most probable decay corresponds to that which produces the daughter nuclei with magic neutron number (N=82). This observation stresses the role of magicity or shell closure in cluster decay process.
We have also plotted the Geiger-Nuttel plot for all the considered decay modes.  The linear nature of the graph is successfully reproduced.  In order to understand the dependence of Skyrme forces in predicting the half-lives, we have to do global calculation throughout the nuclear chart.

\section*{Acknowledgements}

One of the authors, (NA) gratefully acknowledges UGC, Govt. of India, for providing the grant under UGC-JRF/SRF scheme.

\end{document}